# Thulium environment in a silica doped optical fibre

W. Blanc[1], T.L. Sebastian[2], B. Dussardier[1], C. Michel[1], B. Faure[1], M. Ude[1], G. Monnom[1]

[1]*Laboratoire de Physique de la Matière Condensée, CNRS UMR6622 Université de Nice - Sophia Antipolis, Parc Valrose, 06108 Nice Cedex 2, France*

[2]*Departamento de Fisica, Universidad de Concepcion, Casilla – 160 C, Concepcion, Octava Region, Chile*

**Abstract**

Thulium-doped optical fibre amplifiers (TDFA) are developed to extend the optical telecommunication Wavelength Division Multiplexing (WDM) bandwidth in the so-called S-band (1460-1530 nm). The radiative transition at 1.47 µm ($^3H_4 \rightarrow {}^3F_4$) competes with a non-radiative multi-phonon deexcitation ($^3H_4 \rightarrow {}^3H_5$). The quantum efficiency of the transition of interest is then highly affected by the phonon energy ($E_p$) of the material. For reliability reasons, oxide glasses are preferred but suffer from high phonon energy. In the case of silica glass, $E_p$ is around 1100 cm$^{-1}$ and quantum efficiency is as low as 2%. To improve it, phonon energy in thulium environment must be lowered. For that reason, aluminium is added and we explore three different core compositions: pure silica, and silica slightly modified with germanium or phosphorus. The role of aluminium is studied through fluorescence decay curves, fitted according to the continuous function decay analysis. From this analysis, modification of the thulium local environment due to aluminium is evidenced.

# 1. Introduction

Actual long-haul optical telecommunication systems are using repeaters, all based on erbium-doped fibre amplifiers (EDFA) which cover the so-called C- and L-bands (1530-1560 nm and 1570-1610 nm, respectively). As the information traffic in wavelength-division-multiplexing (WDM) optical communication systems is increasing rapidly, it is necessary to extend the telecommunication wavelength range and to develop amplifiers in the S-band (1460-1530 nm). One of the possible solutions is the thulium-doped fibre amplifier (TDFA). Indeed, in thulium an emission occurs at 1.47 µm between the two excited levels $^3H_4$ and $^3F_4$ [1]. This emission perfectly matches the S-band. However, a non-radiative emission occurs from the $^3H_4$ level to the $^3H_5$ level directly located below the emitting level. The probability of this transition depends exponentially on the energy gap and phonon energy. In ZBLAN, phonon energy is about 580 cm$^{-1}$. As the energy gap is around 4000 cm$^{-1}$, more than 6 phonons are required to bridge the $^3H_4$-$^3H_5$ gap. This renders the non-radiative de-excitation very improbable and consequently the efficiency of the 1.47 µm is close to 100% (the measured lifetime is close to the radiative one). But fluoride glass fibres are difficult to fabricate, have some problem of reliability and are hygroscopic. Furthermore, the connection to a standard telecommunication fibre necessitates special care. For these reasons, more robust oxide glasses are preferred. For example, Aitken *et al.* investigated aluminates because of the low phonon energy of this material (780 cm$^{-1}$) [2]. For such glass, quantum efficiency reaches ≈35%. Development of new glasses focuses on lowering the maximum phonon energy (MPE). However, we observed an improvement of the quantum efficiency without lowering this MPE [3]. The glass under investigation was silica. It is usually considered as a bad glass for the application considered here due to it high phonon energy (1100 cm$^{-1}$) yielding at radiative efficiency of the 1.47 µm emission as low as 2%. The improvement of the quantum efficiency was correlated with aluminium (whose $Al_2O_3$ phonon energy is 870 cm$^{-1}$ [4]) without explaining the role of the oxidized glass network modifiers.

In this article, three different high MPE glasses are investigated: pure silica (Si, MPE=1100 cm$^{-1}$), silica slightly modified by germanium (Ge-Si, MPE=1100 cm$^{-1}$) or phosphorus (P-Si, MPE=1300 cm$^{-1}$). Aluminium is incorporated in different concentrations and it influence is studied through thulium decay curves measurements. Their analyses evidence the role of the local phonon energy environment.

## 2. Experimental Details

The fibres investigated were drawn from performs prepared by the Modified Chemical Vapor Deposition at Laboratoire de Physique de la Matière Condensée (Nice). Germanium and Phosphorus were incorporated during the core layer deposition. Germanium concentration was measured to be 4 mol% by Electron Probe Microscopy Analysis. The phosphorus content was estimated to be 1 mol%. Thulium and Aluminium ions were incorporated through the solution doping technique [5]. $Tm^{3+}$ concentration was estimated by measuring the absorption peak at 780 nm and using an absorption cross-section value of $8.7.10^{-25}$ $m^2$ [6]. The concentration in $at/m^3$ was converted into ppm mol assuming that the silica density is 2.2. It was measured to be around 50 ppm mol for all fibres investigated. $Al_2O_3$ concentration was varied from 1.4 to 9 mol%.

$^3H_4$ level lifetime can be measured by monitoring either the 1470 or the 800 nm emission. The last one was chosen to measure luminescence kinetics because it is ten times more efficient considering the branching ratio. The beam from a modulated 785-nm, 20-mW fibre-coupled laser diode was coupled into the tested fibre through a single-mode fibre-coupler. The pump wavelength was tuned to that of the $^3H_6 => {}^3H_4$ absorption peak. The coupler transmitted 60% of the pump to the Tm-doped fibre. To avoid spectral distorsion caused by re-absorption of the signal, the 810-nm fluorescence was measured counterpropagatively. Although a weak residual absorption exists, the setup and procedure were chosen to minimize its impact. The fluorescence was collected from the second coupler arm and 48% of it was directed to an optical spectrum analyser (Anritsu MS9030A-MS9701B) equiped with a –20 dB optical through-put port. This port was tuned to 810 nm (1 nm spectral width). The output light was detected by an amplified avalanche silicon photodiode (APD) (EG&G SPCM AQR-14-FC) operated in the photon-counting mode. The TTL electrical pulses from the APD were counted by a Stanford SR400 photon counter synchronized by the laser diode modulation signal. Decay curves were registered using a time-gate scanning across one pump modulation period. In order to minimize errors (laser fluctuations, …) and increase the S/N ratio, the signal was normalized in real time by the signal from a fixed time-gate integrating the signal over a full 500 μs-modulation period, and each data point was averaged 1000 times. A decay curve contained typically 800 data points.

## 3. Results

The decay curves of the $^3H_4 \rightarrow ^3H_6$ Tm-emission are shown on Fig. 1 for Ge-Si (a), P-Si (b) and Si (c) core compositions for various $Al_2O_3$ concentrations. For the three starting glasses compositions, the same behaviour is observed: when aluminium concentration increases, fluorescence intensity increases at a given time and the shape tends to be more and more exponential. It shows that thulium luminescence kinetics are highly influenced by the aluminium concentration. Decay curves for the three core compositions, with the same aluminium concentration, are represented on Fig. 2. The role of the glass composition seems to be negligible. While the MPE of the three glasses are different, the three decay curves are all the same. The P-Si composition has the highest MPE. Then, a faster luminescence kinetic would be expected. However, aluminium oxide phonon energy is 870 cm$^{-1}$, much lower than the MPE of the glasses. So decay curves from Fig. 2 tend to prove that thulium are sensitive to the local phonon energy and not to the MPE. Elsewhere, it is interesting to note that germanium oxide phonon energy is around 900 cm$^{-1}$, close to the one of aluminium oxide [7]. From the considerations mentioned above, this element should contribute to improve luminescence kinetics. We prepared a thulium-doped silica fibre with 20 mol% of $GeO_2$ and no aluminium. To compare this luminescence kinetic with the Al-doped fibre, we considered the time when the intensity value is 1/e of it initial value. Decay constants obtained are 28 ± 3 μs for the 20 mol % $GeO_2$-doped and 36 ± 3 μs for a 4 mol% $Al_2O_3$-doped silica fibre, respectively. It means that aluminium is more influent than Ge on the thulium environment. This could be explained by the role of these elements. Silica glasses have a covalently bonded structure, implying that a certain concentration of nonbridging oxygen groups must be present in the glass to allow for the incorporation of rare-earth ions. In pure silica glasses, a rigid structure exists and, accordingly, there is a lack of nonbridging Si-O$^-$ groups. This makes the coordination of $Tm^{3+}$ difficult and contributes to the lack of its solubility. On one hand, germanium ion is a former network. Then, when it is inserted into the silica glass, the network is almost the same and thulium has no specific reason to be located close to germanium or silica ions. On the other hand, aluminium ions can be inserted in the $SiO_2$ network as a network modifier element. In this case, the aluminium ions break the covalently bonded tetrahedral structure of $SiO_2$ and produce the nonbridging Al-O$^-$ oxygen group. Then rare-earth ions is preferentially gather near aluminium sites [8,9].

## 4. Discussion

All the measured decay curves have a non-exponential shape. Such behaviour is usually associated with an energy transfer. In the case of thulium, two energy transfers can occur: among thulium ions or from a thulium ion to an OH radical. We demonstrate first that none of these phenomena caused non-exponential decay in our samples.

When excited at 785 nm, thulium-thulium energy transfer occurs through the cross-relaxation process $^3H_4, ^3H_6 \rightarrow ^3F_4, ^3F_4$ and was observed previously by Jackson [10]. An increasing of the cross relaxation rate tends to shorten the $^3H_4$ lifetime. Two fibres were prepared with the same concentration of aluminium (5 mol%) and with three different thulium concentration: 200, 550 and 2900 ppm mol. Under 786 nm excitation pumping scheme, the decay curves of the $^3H_4 \rightarrow ^3H_6$ transition are exactly the same for all the three samples [11]. As thulium concentration is below 100 ppm for all the fibres studied here, we can not consider that changes in the shape are due to thulium-thulium energy transfer. The second energy transfer mechanism invoked is due to the presence of OH-radicals. This group has an absorption band around 1.38 μm which partially overlaps the 1.47 μm emission band ($^3H_4 \rightarrow ^3F_4$ transition). The OH concentration was estimated by measuring absorption curve in fibres with two different $Al_2O_3$ concentrations: 2.5 and 9 mol%. The same amount was obtained in both cases, 2 ppm. Firstly, it is much lower than the thulium one, therefore the Tm-OH energy transfer lies a low probability. Secondly, as thulium and OH concentrations are the same in fibres, this mechanism has the same probability and can not explained the variations of the decay curves.

The two energy transfer mechanisms invoked can not be used to interpret the shape evolution of the decay curves when Al concentration increases. To explain it we consider that thulium ions are inserted in a glass which is characterized by a multitude of different sites available for the rare-earth ion, leading to a multitude of decay constants. This phenomenological model was first proposed by Grinberg *et al.* and applied to chromium in glasses [12]. Here we apply this model, for the first time to our knowledge, to Tm-doped glass fibres. In this method, a continuous distribution of lifetime rather than a number of discrete contributions is used. The advantage of this method is that no luminescence decay model or physical model of the material is required *a priori*. Assuming that sites are characterized by a continuous distribution of decay constant A($\tau$), the luminescence decay is given by:

$$I(t)=\int \frac{A(\tau)}{\tau}\exp(-t/\tau)d\tau \tag{1}$$

In the calculations, the continuous decay time distribution is replaced by a discrete distribution of logarithmically spaced decay curves:

$$\int A(\tau)\exp(-t/\tau)d(\ln\tau) \approx \sum_i A_i \exp(-t/\tau_i) \tag{2}$$

To recover distribution function A($\tau$) from the experimental luminescence decay, the $\chi^2$ function is minimized. $\chi^2$ is defined as:

$$\chi^2 = \sum_k \frac{[I_{\exp}(t_k) - I(t_k)]^2}{\sigma_k^2} \tag{3}$$

where $I_{ex}(t_k)$ is the experimental emission intensity at time $t_k$, $I(t_k)$ is the calculated intensity given by expression (2), and $\sigma_k$ is the weighting of the data point. The procedure for calculating $\sigma_k$ and the fitting algorithm is described in detail in [12].

For the fitting procedure, we considered 125 different values for $\tau_i$, logarithmically spaced from 1 to 1000 µs. By applying this procedure to all the decay curves presented on Fig. 1, a good matching was generally obtained (Fig. 3), except for some of the decay curves, although apparently very similar to those fitted with success. One can suspect differences in the signal to noise ratio from curve to curve to be at the origin of this problem. To minimize it, it is possible to average data (as made by Grinberg *et al.*[12]) but there is a risk to modify the shape of the curve, especially the beginning part. That is why we preferred to apply the fitting procedure to raw data.

Here we present the most representative fits results. Histograms corresponding to the recovered luminescence decay time distributions obtained for Ge-Si and P-Si core compositions with different Al concentrations are shown on Fig. 4 and 5. For a given composition (Fig. 4), we can notice two main distributions of the decay constant. With the aluminium concentration, they increase from 6 to 15 µs and from 20 to 50 µs. For the highest aluminium concentration (Fig. 5), these two bands are still present. One is around 10 µs and the second one spreads from 30 to 100 µs, for both compositions (Ge-Si and P-Si). According to the phenomenological model, the width of the decay constants distribution is related to the number of different sites. The large distribution around 80 µs is then due to a large number of sites available with different environments. It is however remarkable that this distribution at ≈80 µs is very similar for the two different compositions. From the thulium point of view (considering luminescence kinetics), Ge-Si and P-Si glasses seems to offer the same sites.

The meaning of the decay constant values is now discussed. Lifetime constants obtained from the fitting can be correlated with the one expected for thulium located in an pure silica or pure $Al_2O_3$ environment. The $^3H_4$ lifetime is calculating by using this equation:

$$\frac{1}{\tau}=\frac{1}{\tau_{rad}}+W_{nr} \tag{4}$$

where $\tau_{rad}$ corresponds to the radiative lifetime which is given to be 670 µs in silica [13]. $W_{nr}$ is the non-radiative decay rate, expressed as [14]:

$$W_{nr}=W_0\times\exp\{-\alpha(\Delta E-2E_p)\} \tag{5}$$

where $W_0$ and $\alpha$ are constants depending on the material, $\Delta E$ is the energy difference between the $^3H_4$ and $^3H_5$ levels and $E_p$ is the phonon energy of the glass.

$W_0$ and $\alpha$ were estimated for different oxide glasses [14,15]. Comparison of these values indicates that they are quite similar and even if no data can be found for aluminate glass, we can consider in first approximation the silica values to calculate the lifetime $\tau$. The energy difference $\Delta E$ was estimated by measuring the absorption spectrum of the fibres. When Al concentration varies, this value is almost constant around 3700 cm$^{-1}$. It is slightly lower than the usual 4000 cm$^{-1}$ value given in literature because we did not considered the peak-to-peak difference but the energy difference between the long and short wavelength parts of the $^3H_4$ and $^3H_5$ absorption bands, respectively.

With these considerations, the $^3H_4$ expected lifetime can be calculated. In the case of silica glass, $\tau_{silica}$ = 6 µs and for an $Al_2O_3$ environment, $\tau_{alumina}$ = 110 µs. These two values are in accordance with the ones we obtained from the fitting procedure (Fig. 4). The distribution of decay constant around 10 µs corresponds to thulium ions located in almost pure silica environment while the second distribution of decay constant is attributed to $Tm^{3+}$ located in aluminium-rich sites.

## 5. Conclusion

Thulium-doped fibres amplifiers are proposed to extend the telecommunication network capacity. The quantum efficiency of the 1.47 μm emission is affected by a non-radiative deexcitation. So, the proposed glasses are developed with the lowest maximum phonon energy. In this article, through the continuous function decay analysis, we demonstrate that it is necessary to consider the local phonon energy in the thulium environment. It is then possible to highly improve the quantum efficiency by incorporating aluminium ions while the highest phonon energy remains unchanged.

**Acknowledgements**

The authors gratefully acknowledge Ivan Kasik form the Institute of Radio Engineering and Electronics in Prague, Czech Republic, for EPMA measurements. The post-doc position of TLS was funded by the Ministère de l'Education Nationale, de l'Enseignement Supérieur et de la Recherche, in France.

**References**

[1] B.M. Antipenko, A.A. Mak, O.B. Raba, K.B. Seiranyan, T.V. Uvarova, Soviet Journal of Quantum Electronics 13 (1983) 558.

[2] B.G. Aitken, M.L. Powley, R.M. Morena, B.Z. Hanson, J. Non-Cryst. Solids 352 (2006) 488.

[3] B. Faure, W. Blanc, B. Dussardier, G. Monnom, P. Peterka, Technical Digest of Optical Amplifiers and Their Applications, San Francisco, 2004, OWC2.

[4] M.B. Lee, J.H. Lee, B.G. Frederick, N.V. Richardson, Surf. Sci. 448 (2000) L207.

[5] J.E Townsend, S.B. Poole, D.N. Payne, Elec. Lett. 23 (1987) 329

[6] D.C. Hanna, I.R. Perry, J.R. Lincoln, J.E. Townsend, Opt. Commun. 80 (1990) 52.

[7] C.B. Layne, W.H. Lowdermilk, M.J. Weber, Phys. Rev. B. 16 (1977) 10.

[8] Y. Zhou, Y.L. Lam, S.S Wang, H.L. Liu, C.H. Kam, Y.C. Chan, Appl. Phys. Lett. 71 (1997) 587.

[9] A. Monteil, S. Chaussedent, G. Alombert-Goget, N. Gaumer, J. Obriot, S.J.L. Ribeiro, Y. Messaddeq, A. Chiasera, M. Ferrari, J. Non Cryst. Solids 348 (2004) 44.

[10] S.D. Jackson, Opt. Commun. 230 (2004) 197.

[11] D.A. Simpson, G.W. Baxter, S.F. Collins, W.E.K Gibbs, W. Blanc, B. Dussardier, G. Monnom, J. Non-Cryst. Solids 352 (2006) 136.

[12] M. Grinberg, D.L. Russell, K. Holliday, K. Wisniewski, Cz. Koepke, Opt. Commun., 156 (1998) 409.

[13] B.M. Walsh, N.P. Barnes, Appl. Phys. B 78 (2004) 325.

[14] J.M.F van Dijk, M.F.H. Schuurmans, J. Chem. Phys. 78 (1983) 5317.

[15] C.B. Lain, Phys. Rev. B 16 (1977) 10.

**List of Figures:**

**Figure 1**

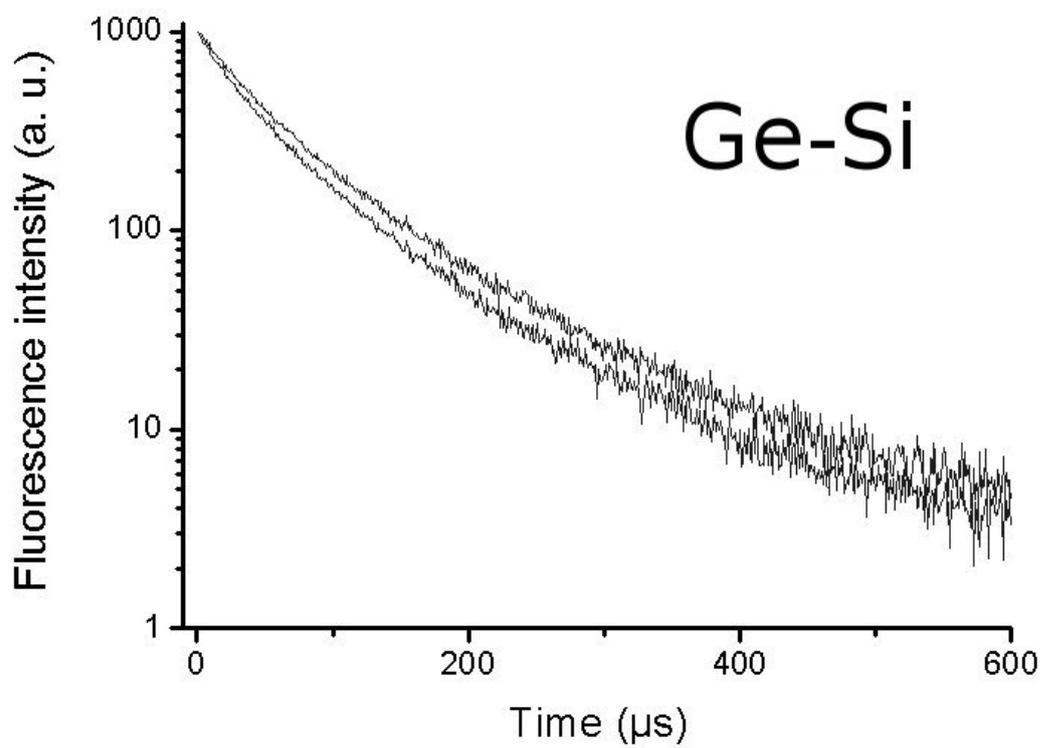

(a)

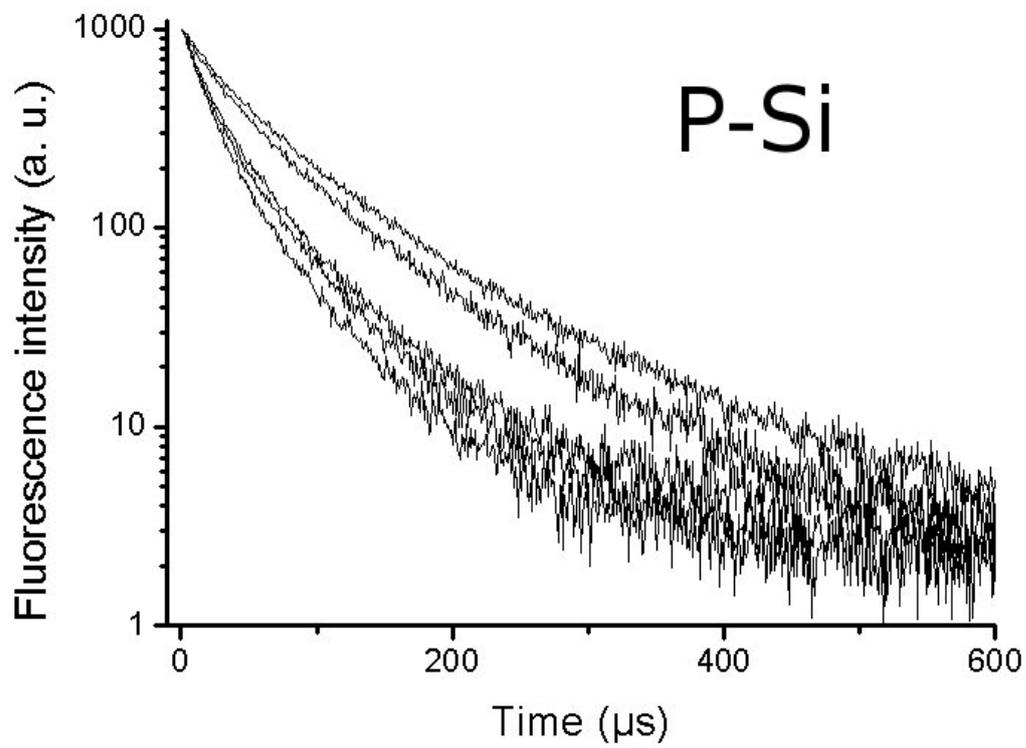

(b)

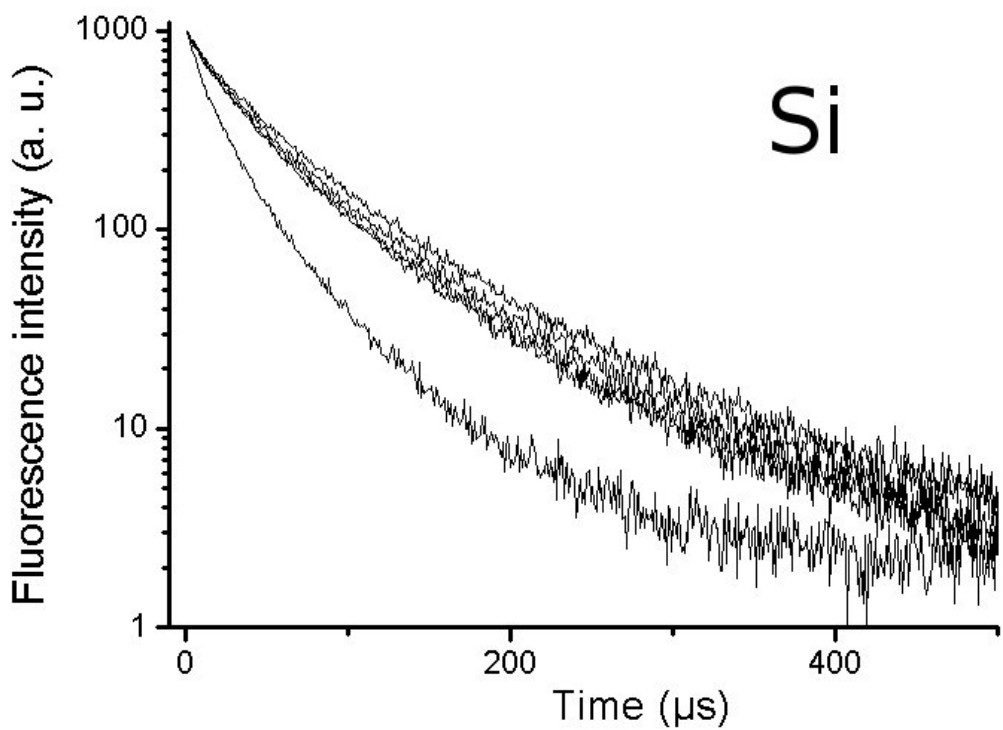

(c)

**Figure 2**

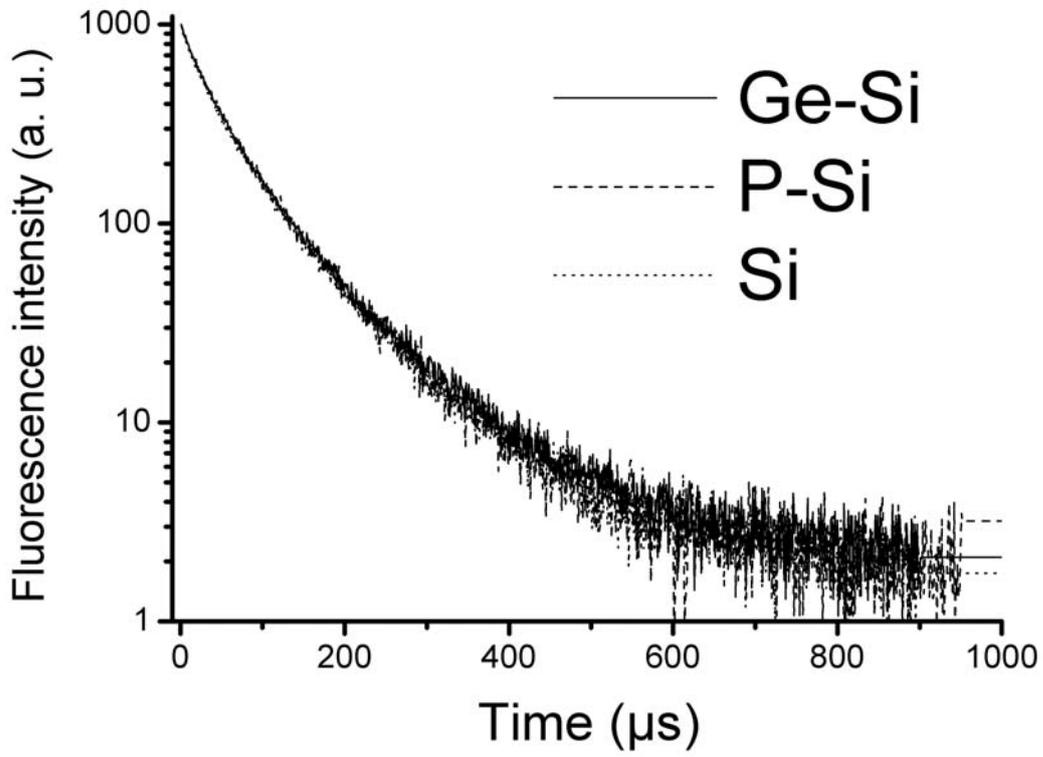

**Figure 3**

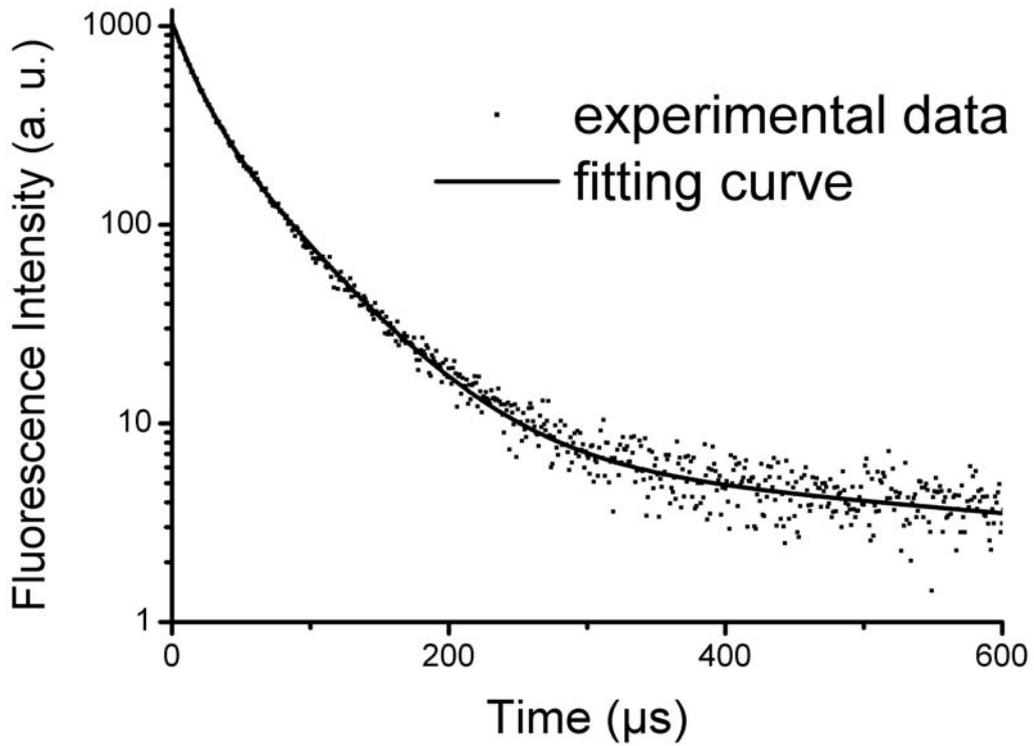

**Figure 4**

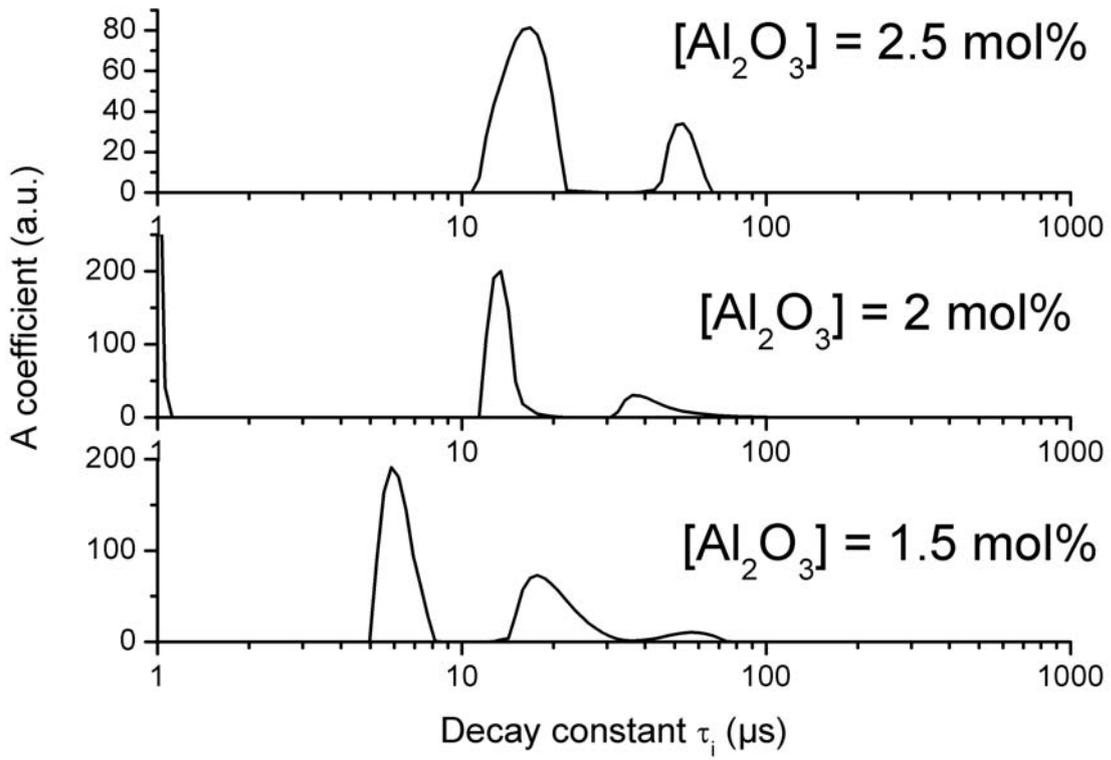

**Figure 5**

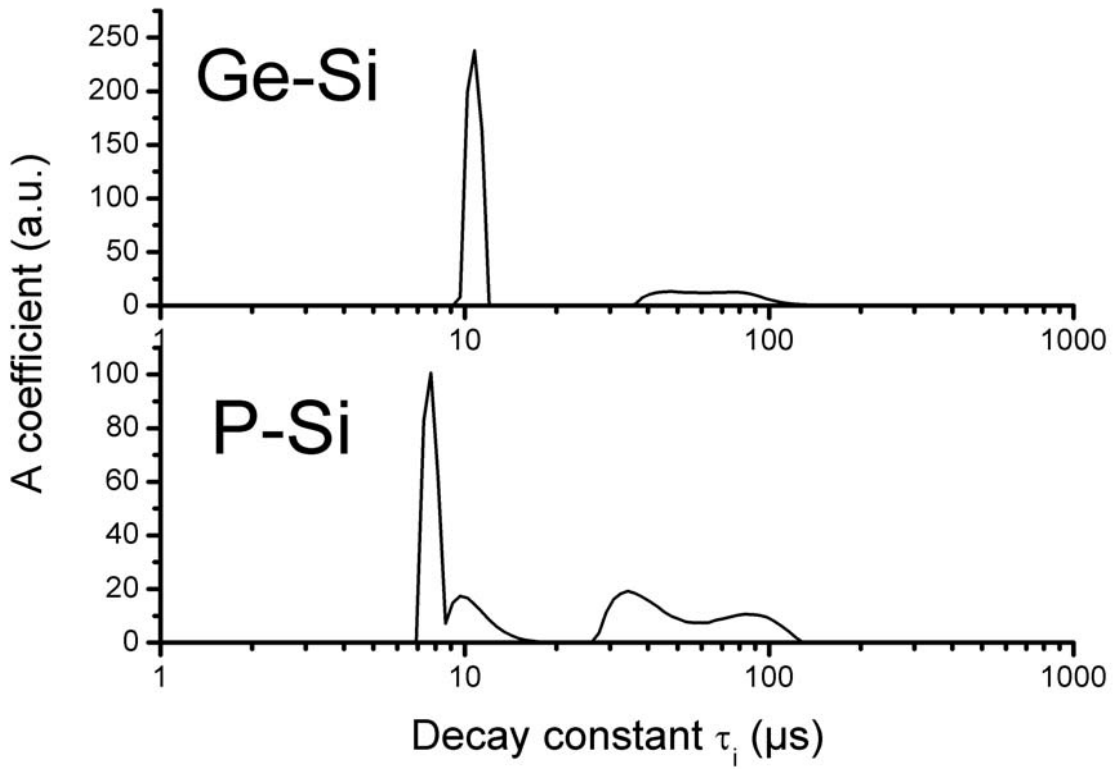

**Figures Captions**

Figure 1: luminescence decay curves monitored at 805 nm in silica-based thulium-doped fibres with germanium co-doped (a), phosphorus co-doped silica (b) or pure silica (c) core composition. $Al_2O_3$ content is (from the lowest curve to the highest) 7 and 9 mol% (a), 1.5, 2, 2.5, 7 and 9 mol% (b) and 1.4, 4, 5, 6, 7 mol% (c).

Figure 2: luminescence decay curves monitored at 805 nm in silica-based thulium-doped fibres with different core compositions. $Al_2O_3$ concentration is 7 mol%.

Figure 3: luminescence decay curve monitored at 805 nm in silica-based thulium-doped fibres with phosphorus and 2.5 mol% $Al_2O_3$ in the core (dotted line). The solid curve represents the fit obtained from recovered lifetime distributions.

Figure 4: histograms of the recovered luminescence decay time distributions obtained for silica-based thulium-doped fibres with phosphorus incorporated in the core and different $Al_2O_3$ concentration.

Figure 5: histograms of the recovered luminescence decay time distributions obtained for two compositions of silica-based thulium-doped fibres with the same $Al_2O_3$ concentration of 9 mol%.